\documentclass[useAMS]{mn2e}
\usepackage[tight]{subfigure}
\usepackage{longtable}
\usepackage{amssymb}
\usepackage{lscape}
\usepackage{graphicx}

\title[The UV-upturn in BCGs]{The UV-upturn in brightest cluster galaxies}
\author[Loubser $\&$ S\'{a}nchez-Bl\'{a}zquez]{S. I. Loubser$^{1}$\thanks{E-mail:
Ilani.Loubser@nwu.ac.za (SIL)}, P. S\'{a}nchez-Bl\'{a}zquez$^{2,3,4}$\\
$^{1}$Unit for Space Physics, North-West University, Potchefstroom 2520, South Africa\\
$^{2}$Instituto de Astrof\'{\i}sica de Canarias, E-38200 La Laguna, Tenerife, Spain\\
$^{3}$Departamento de Astrof\'{\i}sica, Universidad de La Laguna, E-38205 La Laguna, Tenerife, Spain\\
$^{4}$Departamento de F\'{\i}sica Te\'orica, M\'odulo C15, Universidad Aut\'onoma de Madrid, E28049, Cantoblanco, Spain}
\begin{document}


\pagerange{\pageref{firstpage}--\pageref{lastpage}} \pubyear{2010}

\maketitle

\label{firstpage}

\begin{abstract}
This paper is part of a series devoted to the investigation of a large sample of brightest cluster galaxies (BCGs), their properties and the relationships between these and the properties of the host clusters. In this paper, we compare the stellar population properties derived from high signal-to-noise, optical long-slit spectra with the $GALEX$ ultraviolet (UV) colour measurements for 36 nearby BCGs to understand the diversity in the most rapidly evolving feature in old stellar systems, the UV-upturn. 
We investigate: (1) the possible differences between the UV-upturn of BCGs and those of a control sample of ordinary ellipticals in the same mass range, as well as possible correlations between the UV-upturn and other general properties of the galaxies; (2) possible correlations between the UV-upturn and the properties of the host clusters; (3) recently proposed scenarios where helium-sedimentation in the cluster centre can produce an enhanced UV-upturn. We find systematic differences between the UV-colours of BCGs and ordinary ellipticals, but we do not find correlations between these colours and the properties of the host clusters. Furthermore, the observations do not support the predictions made by the helium-sedimentation model as an enhancer of the UV-upturn.
\end{abstract} 


\section{Introduction}
\label{Introduction}

The UV-upturn, first discovered by Code $\&$ Welch (1979), is the rising flux with decreasing wavelength from 2500 \AA{} to the Lyman limit observed in the spectrum of some massive elliptical galaxies (see O'Connell 1999 for a review).

In spiral and irregular galaxies, flux emitted in the UV acts as an excellent measure of the current star formation rate (Gil de Paz et al.\ 2007). In the case of (mostly quiescent) early-type galaxies, the flux at wavelengths shorter than 1800 \AA{} (FUV) is believed to be produced by hot, low mass Horizontal Branch (HB) stars, although the mechanism with which an old population can develop hot stars is still hotly debated (see Yi 2008). For the non-quiescent, low redshift elliptical galaxies, analysis of UV wavelengths longer than 1800 \AA{} (NUV) is a useful indicator of residual star formation (Pipino et al.\ 2009). Possible solutions for the origin of the increase in FUV flux in quiescent early-type galaxies include metal rich stars that lose mass and become HB stars, metal poor stars that evolve to very high temperatures and onto the HB, or mass transfer in binary systems that can expose the hot core of a HB star (Donahue et al.\ 2010). These theoretical scenarios make different predictions about the correlations between the strength of the UV-upturn and the metallicity, as well as about the evolution of this feature with redshift in passively evolving systems. Observations of the evolution of the FUV--V colours of Brightest Cluster Galaxies (BCGs) with redshift seem to favour the single stars hypothesis above the binary theory (Ree et al.\ 2007). However, the metal poor and metal rich hypotheses also have problems. For example, a large fraction of metal poor stars are needed to explain the strength of the UV-upturn observed in elliptical galaxies and the metal rich scenario seems to require mass losses from Red Giant Branch stars much larger than those predicted theoretically.
 
A popular third possibility, that seems to solve some of the problems associated with the HB hypotheses, is the enhanced helium abundance hypothesis. This hypothesis is motivated by globular cluster observations containing peculiar features in the colour magnitude diagram that can only be explained by the presence of super-helium-rich sub-populations (D'Antona $\&$ Caloi 2004; Norris 2004; D'Antona et al. 2005; Lee et al.\ 2005; Piotto et al. 2005). These populations could be a major source of FUV flux in quiescent elliptical galaxies, although studies suggest that the deduced value of the helium abundance is too small to be influential on galactic scales (Lee et al.\ 2005; Yi 2008). However, in the centres of clusters, an extra source of helium could come from sedimentation processes (Peng $\&$ Nagai 2009). If the helium sedimentation mechanism is important, the UV-upturn strength in the BCGs could be greatly enhanced. In fact, it is known from observations of nearby early-type galaxy samples, that BCGs are most likely the strongest UV emitters among the quiescent early-type galaxies in each redshift bin (Burstein et al.\ 1988; Boselli et al.\ 2005; Donas et al.\ 2007).

If the mechanism producing the UV-upturn in BCGs is different than that of ordinary elliptical galaxies, then the results on the evolution of the UV-colours with redshift mentioned above will not necessarily be representative for the whole population of objects hosting old stellar populations. In fact, Atlee, Assef $\&$ Kochanek (2009) obtained very different results for UV-colour evolution using a sample of ordinary bright elliptical galaxies.

In this work, we want to explore the mechanism(s) producing the UV-upturn in BCGs and compare it with that producing the UV-upturn in elliptical galaxies by studying the relation between the UV-upturn strength and other properties of the galaxies. Several authors have found correlations between the strength of the UV-upturn and the Mg-index, an indicator of metallicity, in early-type galaxies (Burstein et al.\ 1988; Donas et al.\ 2007; but also see Rich et al.\ 2005). The question remains whether correlations between the UV-upturn and the spectral properties of BCGs exist, which will provide us with further clues about the origin of the UV-upturn.


This paper is the third in a series of papers investigating an overall sample of 49 BCGs in the nearby Universe for which we have obtained high signal-to-noise ratio, long-slit spectra on the Gemini and WHT telescopes. This large, spatially-resolved, spectroscopic sample of BCGs allows possible connections between the kinematical, dynamical and stellar population properties to be studied. The first paper was devoted to the spatially resolved kinematics of the BCGs (Loubser et al.\ 2008, hereafter Paper 1). In the second paper, we derived Single Stellar Population (SSP)-equivalent ages, metallicities and $\alpha$-abundance ratios in the centres of the galaxies, and then systematically compared these with a sample of ordinary elliptical galaxies, as well as with the host cluster properties (Loubser et al.\ 2009, hereafter Paper 2). 

Here, we turn our attention to the UV-upturn and its correlations with the properties derived from the optical spectra, as well as with the host cluster environment. Our BCG sample has the advantage of having detailed X-ray values of the host cluster properties available from the literature, which BCGs extracted from the SDSS lacks. The spectra in this BCG sample also have significantly improved signal-to-noise ratios compared to previous samples (Crawford et al.\ 1999, and references therein). Most of the previous observational studies of BCGs in the UV concentrate on detecting star formation using the NUV measurements, and in mainly cooling flow clusters (Pipino et al.\ 2009; Donahue et al.\ 2010; Wang et al.\ 2010; Hicks, Mushotzky $\&$ Donahue 2010, but also see O'Dea et al.\ 2010 for FUV measurements). Here, we investigate the variations in the UV-upturn, which in turn can cause scatter in the star formation relations derived. In Section \ref{Data}, we describe the UV and optical data. In Section \ref{Mass}, we compare the UV-upturn measured for the BCGs to that measured for a control sample of ordinary ellipticals matched in mass range and distribution. In Section \ref{Model predictions}, we test the model predictions by Peng $\&$ Nagai (2009), which predicts that the UV-upturn phenomenon should be most pronounced in BCGs hosted by high mass, dynamically relaxed, and cool-core clusters. In Section \ref{Correlations}, we investigate possible correlations between the UV-upturn and the luminosity-weighted age and metallicity for the BCG and ordinary elliptical samples. We summarise our findings in Section \ref{Discussion}.

\section{Data}
\label{Data}

We briefly review the relevant selection criteria and properties of the resulting sample (detailed in Paper 1) here. The sample consists of the dominant galaxies closest to the X-ray peaks in the centres of their host clusters. We adopt the definition to comply with recent literature (for example De Lucia $\&$ Blaizot 2007; Von der Linden et al.\ 2007), where the central, dominant galaxy in a cluster is referred to as the BCG. According to the above definition, for a small fraction of clusters the BCG might not strictly be the brightest galaxy in the cluster. 

We initially intended to investigate a subsample of BCGs with extended haloes (cD galaxies). However, due to the difficulties in the classification of cD galaxies and the very inhomogeneous definitions in the literature, we cannot be confident that all the galaxies in our sample are cD galaxies. Instead, we can say that our sample comprises of the dominant galaxies closest to the X-ray peaks in the centres of clusters. The sample selection combined three methods making the best use of available information from literature and astronomical databases: the two well-known galaxy cluster classification systems that distinguish clusters containing a cD galaxy in the centre from other galaxy clusters (Rood $\&$ Sastry 1971; Bautz $\&$ Morgan 1970); an all-sky search in the HyperLEDA\footnote{http://leda.univ-lyon1.fr/.} database for galaxies with T-types between -3.7 and -4.3 (the de Vaucouleurs Third Reference Catalogue classifies cD galaxies as T-type = --4); and galaxies that broke the de Vaucouleurs $r^{\frac{1}{4}}$ law at large radii. 

In summary: the 49 BCGs in our overall sample were classified as cD either in NED and/or have profiles breaking the $r^{\frac{1}{4}}$ law in the external parts. 
The following global criteria were applied to the sample: apparent B-magnitude brighter than 16; distance closer than 340 Mpc; and an absolute magnitude cutoff at $M_{B}$ = --20. The sample is not complete over any of these parameters, and no UV or X-ray criteria were involved in the selection. Nevertheless, the galaxies are hosted by clusters with a wide range of X-ray luminosities from $L_{\rm X}=0.03 \times 10^{44}$ to $17.07 \times 10^{44}$ erg s$^{-1}$, and it includes BCGs hosted by cooling flow and non-cooling flow clusters.

For this optically-selected sample, high signal-to-noise, long-slit spectra were obtained at the Gemini and WHT telescopes. The spectroscopic data, together with the reduction procedures were presented in Papers 1 and 2, and will not be repeated here. The overall BCG sample has 36 galaxies with FUV and NUV measurements in the $GALEX$ database (ranging from FUV -- NUV = 0.20 to 1.32), which will be analysed further. This subsample of BCGs (still hosted by clusters with a wide range of X-ray luminosities from $L_{\rm X}=0.04 \times 10^{44}$ to $17.07 \times 10^{44}$ erg s$^{-1}$) are representative of our overall BCG sample. Information from $ROSAT$ X-ray data on the properties of the host clusters were collected from the literature (see Section \ref{Model predictions}). Of these 36 BCGs, 18 have X-ray temperature ($T_{\rm X}$) measurements, 27 cluster velocity dispersion ($\sigma_{cluster}$) measurements and 30 X-ray offset (R$_{off}$) measurements. All the galaxy and host cluster properties used here are listed in Table \ref{BigT}.

The $GALEX$ FUV (1344 -- 1786 \AA{}) and NUV (1771 -- 2831 \AA{}) magnitudes and their errors were extracted from the $GALEX$ database\footnote{http://galex.stsci.edu/GR4/} (see Martin et al.\ 2005). Positional cross-correlation of the UV sources in the database was performed within six arcsec of the optical source coordinates, and the UV images were examined by eye to ensure that none of the sources were mismatched, and that the sources were not near the edge of the detector. In each case, the NUV and FUV magnitudes and errors were taken from the longest exposures in the $GALEX$ database, and are mainly from the All-sky Imaging Survey (AIS). The extracted NUV and FUV magnitudes are taken from SExtractor ``mag$\_$auto'' and are the total magnitudes within an elliptical aperture that is scaled to 2.5 times the Kron diameter as determined from the NUV and FUV images, respectively (Kron 1980; Bertin $\&$ Arnouts 1996). To minimise aperture effects and instrumental inhomogeneities, we choose to characterise the UV-upturn as the FUV--NUV colour, based on $GALEX$ measurements alone. In addition, there is a lack of enough homogeneous V (or r) magnitudes available in the literature to keep our sample statistically significant if we characterise the UV-upturn as FUV--V. We recalculated the FUV magnitudes so that the effective FUV apertures were equal to the NUV apertures, by using the flux measurements from the $GALEX$ database and the AB magnitude system of Oke $\&$ Gunn (1983). The $GALEX$ database defines the NUV and FUV magnitudes as $m_{UV}=m_{0}-2.5\log f_{UV}$ where $f_{UV}$ is the count rate and $m_{0}=20.082$ (NUV) and 18.817 (FUV) (Morrissey et al.\ 2005). The uncertainty in these zero points is $\pm$0.15 mag for both FUV and NUV (Donas et al.\ 2007). We propagate the errors on the magnitudes, extracted from $GALEX$, throughout the study.

Internal extinction is small in most nearby early-type galaxies, and the FUV--NUV colour is essentially reddening free for moderate amounts of reddening. We expect a very low impact of the internal extinction on the scatter of the measurements, and as these corrections are very uncertain, we do not include them.

To quantify the effect of Galactic reddening, the Galactic extinction was inferred from the 100$\mu$m dust emission maps of Schlegel, Finkbeiner $\&$ Davis (1998). Using the extinction law of Cardelli, Clayton $\&$ Mathis (1989), Wyder et al.\ (2007) calculated $\frac{A_{FUV}}{E(B-V)}=8.24$ and $\frac{A_{NUV}}{E(B-V)}=8.20$, which gives corrections in the FUV--NUV colours of $\ll$ 0.01 mag, for the E(B--V) values derived for the galaxies in our sample\footnote{When different values such as $\frac{A_{FUV}}{E(B-V)}=8.24$ and $\frac{A_{NUV}}{E(B-V)}=8.10$ (Atlee et al.\ 2009) are used, the corrections remain negligible.}. Hence, these corrections are negligible and not included here. Similarly, at the low redshift of this sample (mean $z=0.0374\pm0.0029$), the $K$-corrections to the magnitudes are negligible.

\begin{figure}
   \centering
   \includegraphics[scale=0.5]{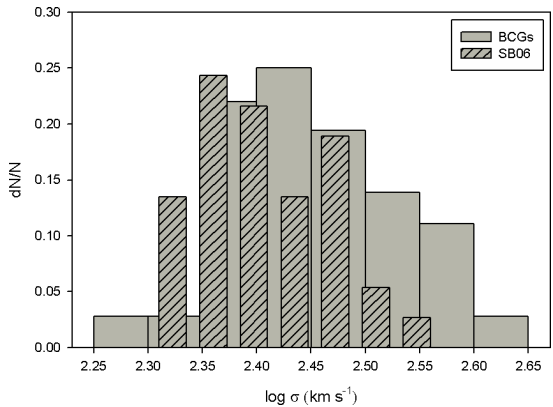}
   \caption[The velocity dispersion distributions of the SB06 and the BCG subsamples with UV measurements.]{The velocity dispersion distributions of the SB06 (grey -- striped) and the BCG (grey -- solid) subsamples with UV measurements.}
   \label{fig:Sig_distr}
\end{figure}

One of goals of this study is to compare the behaviour of the UV-upturn in BCGs with a control sample of ordinary elliptical galaxies. Stellar populations in early-type galaxies seem to be strongly correlated with the central velocity dispersion, and thus we choose to compare the UV-upturn of the BCG sample with a control sample of normal elliptical galaxies with the same range of central velocity dispersion (which we use as a proxy for mass). For this control sample of normal elliptical galaxies, we use the sample of S\'{a}nchez-Bl\'{a}zquez et al.\ (2006, hereafter SB06) in the same mass range as the sample of BCGs, and excluding any known BCGs. The complete SB06 sample consists of 98 galaxies, of which 35 belong to the Coma cluster, and the rest are galaxies in the field, in groups or in the Virgo cluster. The SB06 control sample has 37 galaxies with FUV and NUV measurements in the $GALEX$ database. We performed a Kolmogorov--Smirnov (K--S) test on the velocity dispersion distributions of the two subsamples for which UV measurements were available within the $\log \sigma$ = 2.25 to 2.65 km s$^{-1}$ range (shown in Figure \ref{fig:Sig_distr}), where the null hypothesis is that the distributions were drawn from an identical parent population. The two velocity dispersion distributions are consistent, and have a test value of 0.260, where a test value larger than $D=0.290$ indicates that the two samples compared are significantly different from each other at the 95 per cent confidence level. We used the same procedure described above to extract FUV--NUV colours for the control sample.

\begin{figure}
   \centering
   \includegraphics[scale=0.51]{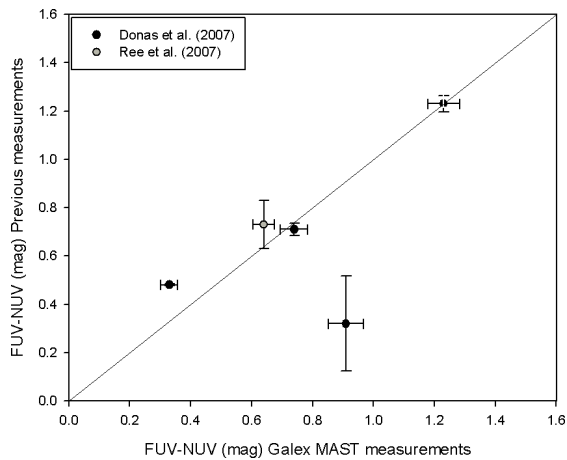}
   \caption{Comparison between the measurements extracted from the $GALEX$ archive against previous measurements by different authors directly from the images.}
   \label{fig:Comparison}
\end{figure}

Five galaxies in our BCG sample have $GALEX$ NUV and FUV measurements which were used in previous studies with similar apertures (Ree et al.\ 2007; Donas et al.\ 2007). In these studies, the magnitudes were extracted directly from the images by the authors. We detect no systematic difference between our FUV--NUV measurements and errors and those of the previous studies plotted in Figure \ref{fig:Comparison}. There is one galaxy (UGC00579) in the Donas et al.\ (2007) sample with a very different colour value. We have investigated this galaxy without finding any reason for the difference, and we have checked that keeping this galaxy included in the sample makes no difference to any of the plots and conclusions. For the rest of the galaxies the values presented here and those measured in the other studies agree within the errors. 

The weak optical emission lines present in some of the BCG spectra most likely indicate active star formation regions or active galactic nuclei (AGN, Edwards et al.\ 2007; Von der Linden et al.\ 2007), both of which can alter the UV properties of the galaxies. For example, there has been some suggestion that relatively tiny amounts of young stars can contribute to the observed scatter in the UV-upturn (Han, Podsiadlowski $\&$ Lynas-Gray 2007; Donas et al.\ 2007). To isolate those galaxies whose UV-spectra may be contaminated by star formation or AGN, we identify the BCGs containing emission lines in their optical spectra, using the \textsc{gandalf} routine\footnote{We make use of the corresponding \textsc{ppxf} and \textsc{gandalf idl} (Interactive Data Language) codes which can be retrieved at http:/www.leidenuniv.nl/sauron/.} (Sarzi et al.\ 2006) as described in Paper 2. The galaxies with detected emission lines will be indicated separately in some of the figures. Of the 36 BCGs studied here, eight have been identified as containing emission lines in their spectra (see Paper 2). 

We measured the individual emission lines and plotted them on a [OIII]$\lambda$5007/H$\beta$ against [OII]$\lambda$3727/H$\beta$ diagram (as [NII]$\lambda$6584 and H$\alpha$ were outside our wavelength range) to separate the two major origins of emission: star formation and AGN (Baldwin, Phillips $\&$ Terlevich 1981; Lamareille et al.\ 2004) in Paper 2. We found all nine galaxies for which the H$\beta$, [OII]$\lambda$3727 and [OIII]$\lambda$5007 lines could be measured (within at least 2$\sigma$ detections) to be star forming galaxies according to this test. Because the diagnostic diagram used in Paper 2 is much less effective at separating different sources of ionisation (Stasi\'{n}ska et al.\ 2006), and because the fraction of the current sample containing emission lines is relatively small, and with weak emission lines, we do not draw detailed conclusions about the nature of emission lines in BCGs here. It can be seen from Figures \ref{fig:nonBCGs}, \ref{fig:Tx}, \ref{fig:Roff} and \ref{fig:Kmags} that including or excluding the emission line BCGs in this study does not influence the conclusions drawn from these plots. It can also be seen in Figure \ref{fig:nonBCGs} that these galaxies are not necessarily bluer in FUV--NUV than the quiescent galaxies. 

Several previous studies which observed BCGs in the NUV (which is sensitive to the detection of young stars), have found a relation between BCGs which show evidence of recent star formation episodes and cooling flows in their host clusters (Mittaz et al.\ 2001; Martel et al.\ 2002; O'Dea et al.\ 2004; Hicks $\&$ Mushotzky 2005; Rafferty et al.\ 2008; Pipino et al.\ 2009; Hicks, Mushotzky $\&$ Donahue 2010). However, not all BCGs in cooling flows show significant star formation (Quillen et al.\ 2008). The UV-upturn is observed in the FUV (which is more sensitive to the HB stars), but as mentioned above, it is important to assess the influence of possible small episodes of recent star formation, as detected in the NUV, on our measurements. Rafferty et al.\ (2008) found that active star formation in BCGs depends on the presence of cooling flows clusters with short cooling times, with the additional requirement of the BCG being located close to the X-ray centre ($\sim20$ kpc). It is therefore also of interest to identify the BCGs hosted by cooling flow clusters, to asses the effects of possible cooling-flow induced star formation in the sample. 

We use different symbols in some of the plots to indicate galaxies hosted by cooling flow clusters, host clusters with definite non-detections of cooling flows, and host clusters for which no cooling flow information could be found. Several definitions for cooling flow clusters exist in the literature, often based on central temperature drop, short cooling time, or significant mass deposition rates. We extract the cooling flow information (see Table \ref{BigT}), based on mass deposition rates, mainly from White, Jones $\&$ Forman (1997), and substitute it with information from Rafferty et al.\ (2006), Edwards et al.\ (2007), and Giovannini, Liuzzo $\&$ Giroletti (2008). The information on the galaxies in common between between these studies agree very well. Of the eight emission line galaxies, two are hosted by cooling flow clusters, three by non-cooling flow clusters, and three for which the cluster cooling flow status is not known. Of the 28 quiescent BCGs, eight are in cooling flow clusters, 11 in non-cooling flow clusters, and nine not known.

The assessment of the impact of star formation on the UV colours is difficult to do, because the main contributors to these colours are still a matter of debate. Stellar population models including post-asymptotic giant branch stars (but not extreme HB stars) indicate that FUV--NUV colours are indeed sensitive to the age of the population, but not more than FUV--V colours, for example. In a population composed of old stars with a small fraction of young stars,  the FUV-NUV colours are affected only when the age of the burst is between 1 -- 2 Gyr (the exact limits depend on the fraction of young stars, see figure 12 in Rampazzo et al.\ 2007). To conclude, we have tested that even if we take only galaxies in non-cooling flow clusters and without emission lines, our conclusions hold.

\section{The UV-upturn as a function of mass in BCGs versus non-BCGs}
\label{Mass}

\begin{figure}
   \centering
   \mbox{\subfigure{\includegraphics[scale=0.53]{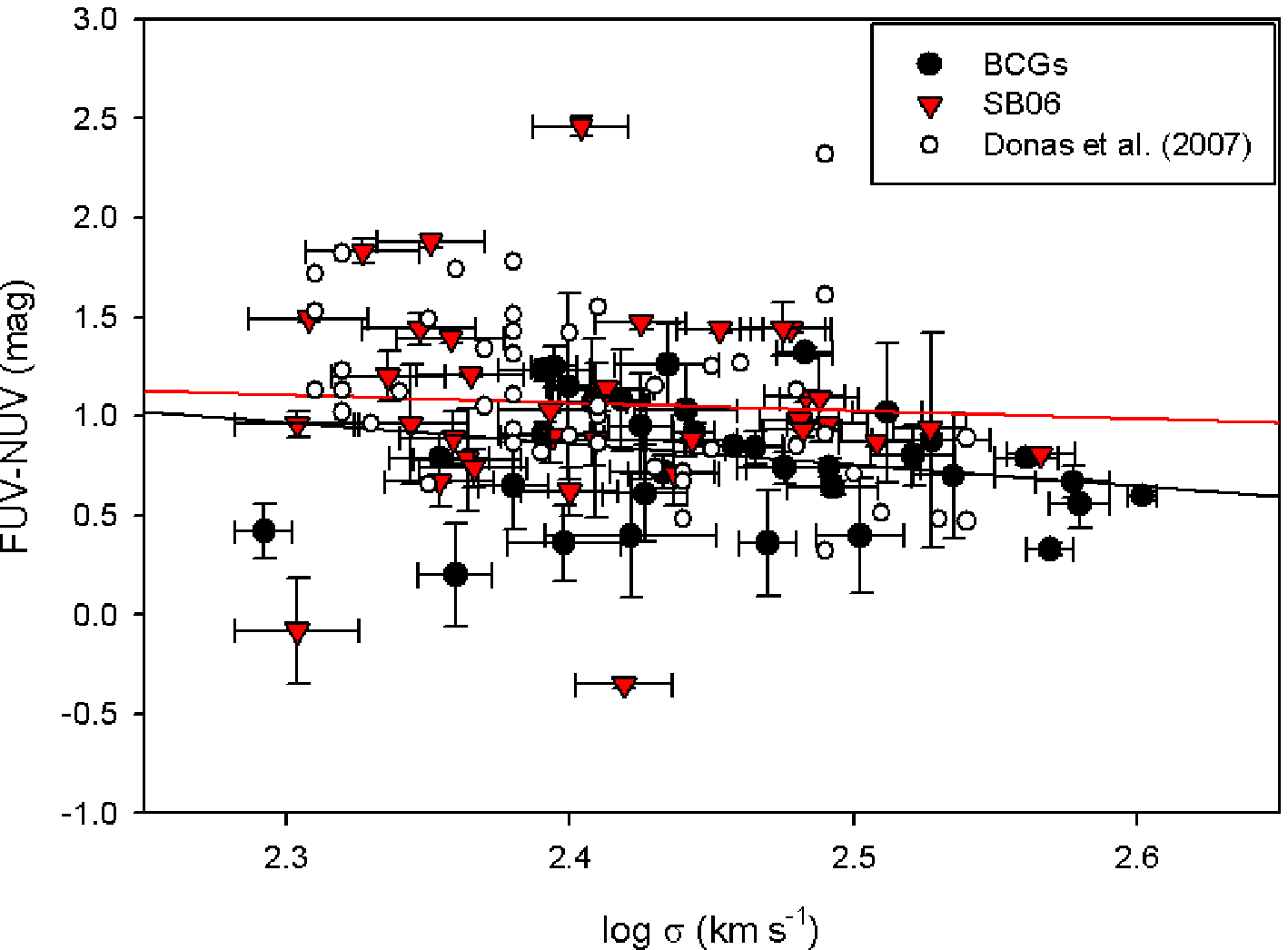}}}
   \mbox{\subfigure{\includegraphics[scale=0.53]{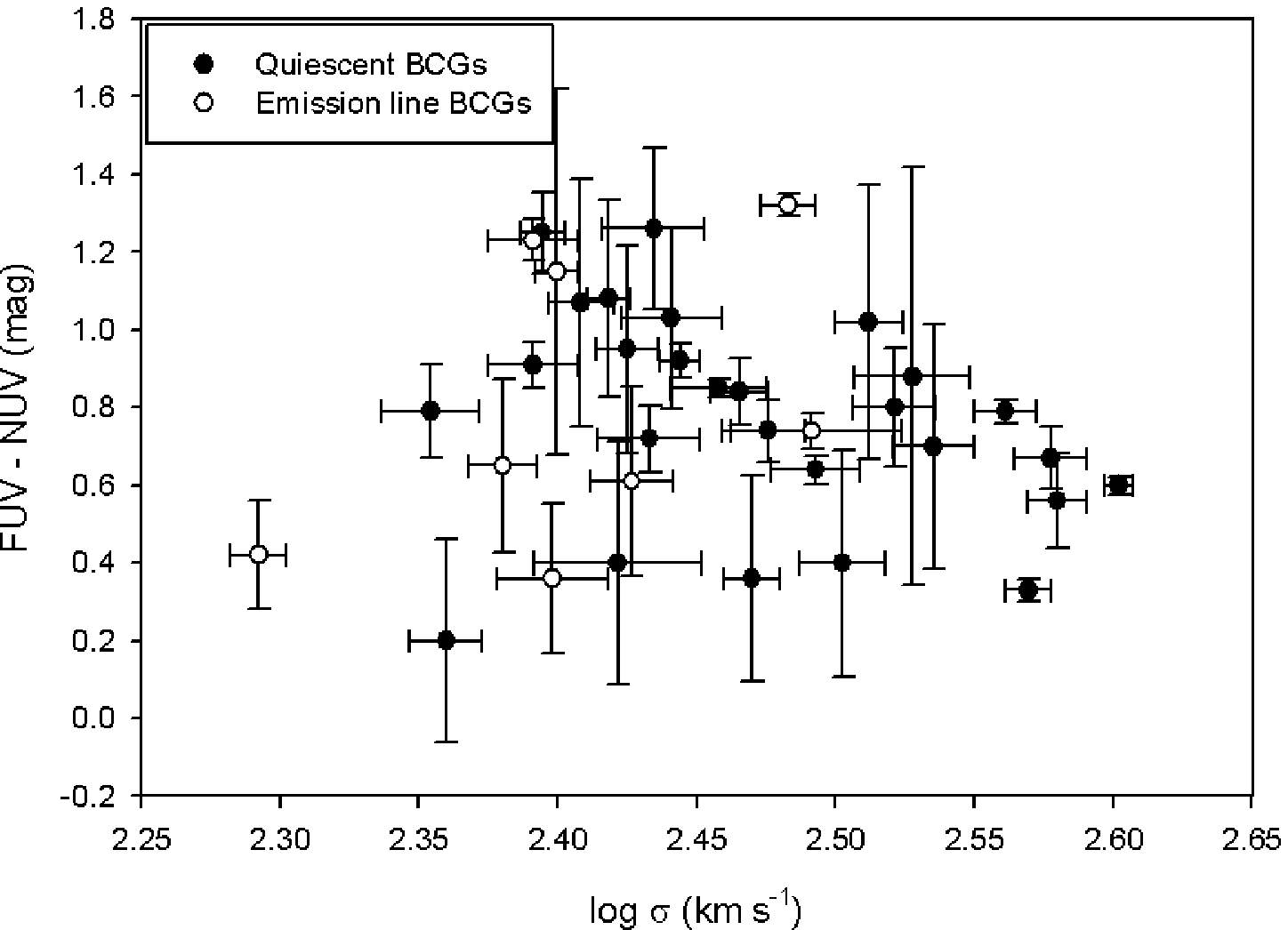}}}
   \mbox{\subfigure{\includegraphics[scale=0.53]{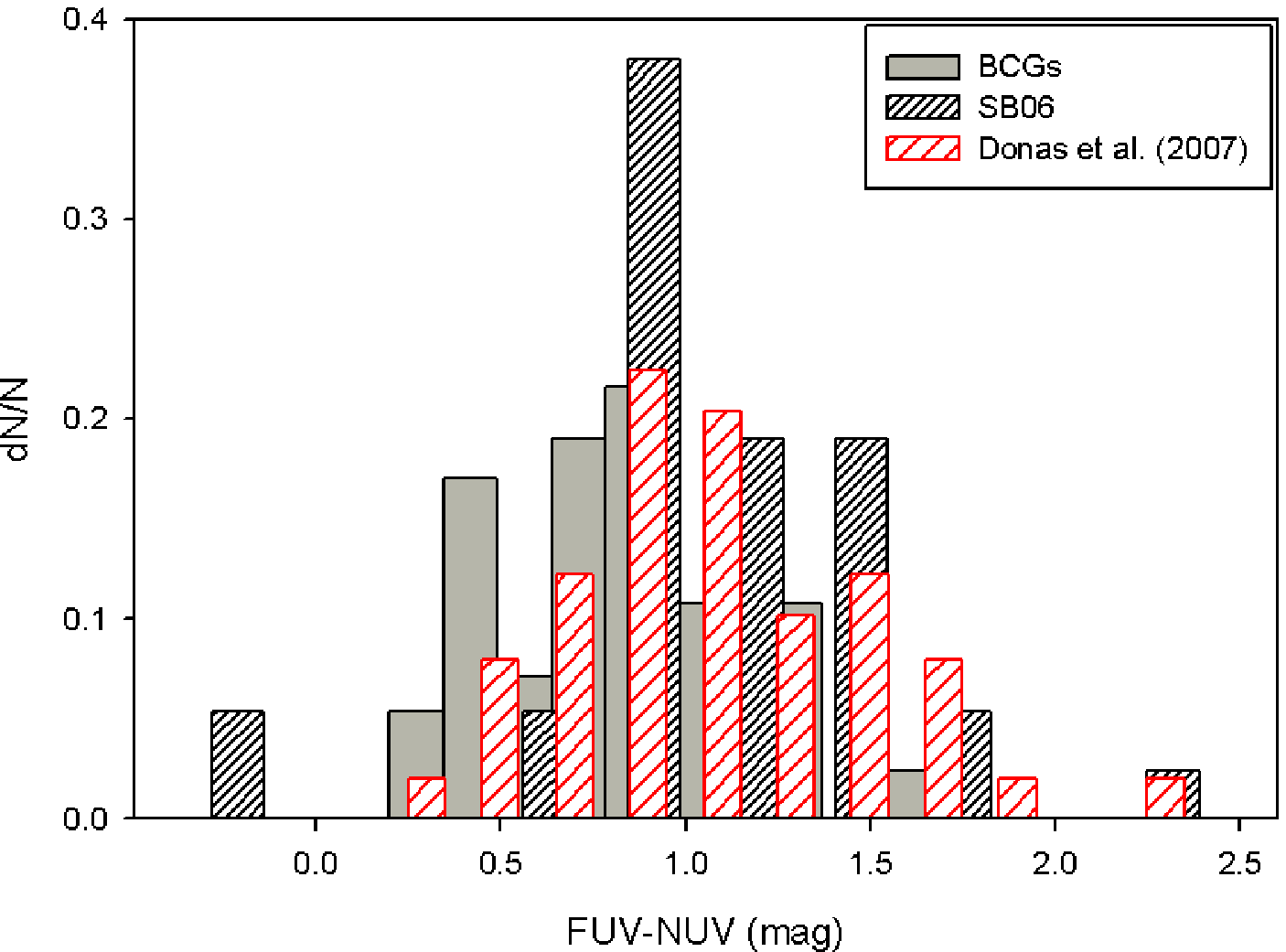}}}
   \caption{The top panel shows the FUV--NUV colour against velocity dispersion for BCGs and non-BCGs. The middle panel shows the FUV--NUV colour against velocity dispersion for just the BCGs, where the filled symbols indicate the quiescent BCGs, and the empty symbols indicate the emission-line BCGs. The bottom panel shows the histogram of the FUV--NUV colour distributions of the BCGs and non-BCGs.}
   \label{fig:nonBCGs}
\end{figure}

Donahue et al.\ (2010) suggested that their sample of BCGs show less dispersion in their NUV--optical colours than that of ellipticals as a whole. They posed the question whether this uniformity in NUV meant that BCGs are less contaminated from star formation or are more uniform in their age and metallicity than their lower mass normal elliptical counterparts. To answer this question, and since stellar populations in early-type galaxies seem to be strongly correlated with the central velocity dispersion, as mentioned, we investigate the UV-upturn (FUV--NUV colour) of the BCG sample relative to the SB06 control sample of ordinary ellipticals in the same mass range. 

We plot the FUV--NUV colours for both the BCG and the control samples in the top panel of Figure \ref{fig:nonBCGs}, and the histograms of the distributions in the bottom panel of Figure \ref{fig:nonBCGs}. Straight line fits to the velocity dispersion plots were made for both samples with a least-squares fitting routine, and are also shown in the top panel of Figure \ref{fig:nonBCGs}. Statistical t-tests were run to explore the presence of correlations. Even though neither of the two samples form a statistically significant correlation with velocity dispersion (mass) at a 95 per cent confidence level, the BCG sample are much less scattered with velocity dispersion than the non-BCGs over the same mass range. The standard deviation on the mean colour values of the BCGs is 0.33 compared to 0.49 for the SB06 sample\footnote{This is the intrinsic scatter since the standard deviations expected from the mean errors on the colour values are negligible.}. The middle panel shows the FUV--NUV colour against velocity dispersion for just the BCGs, where the filled symbols indicate the quiescent BCGs, and the empty symbols indicate the emission-line BCGs. The outliers in this plot are discussed in Section \ref{Correlations}. We do not find a clear distinction between the active and quiescent BCGs, which is in agreement with the findings of Burstein et al.\ (1988) for normal elliptical galaxies.


To eliminate the possibility that the absence of correlations is due to the small range in velocity dispersions (masses), we used the FUV--NUV measurements of ellipticals presented by Donas et al.\ (2007). They found a clear correlation over a larger velocity dispersion range and used the same apertures as used here. We selected all the ellipticals in their sample (which includes some ellipticals in the centres of clusters) in the same velocity dispersion range used here ($\log \sigma$ = 2.30 to 2.65 km s$^{-1}$) and fitted a correlation and t-test as above. In this limited velocity dispersion range, we still find a correlation between FUV--NUV colour and velocity dispersion for their sample.

The histogram in Figure \ref{fig:nonBCGs} indicates another difference in the distributions of the FUV--NUV colours of the BCG and non-BCG samples, with the ordinary ellipticals having larger FUV--NUV colours (i.e.\ having a smaller UV-upturn). We performed a K--S test on the FUV--NUV colour measurement distributions of the two samples, where the null hypothesis is that the distributions were drawn from an identical parent population. We find that the distributions of the two samples compared are significantly different from each other at the 95 per cent confidence level (this has also been confirmed with a chi-squared test). The mean FUV--NUV for BCGs is 0.791$\pm$0.055, whereas it is 1.062$\pm$0.080 for the ellipticals.

To further test the validity of these findings, we also compare the UV colours of the BCGs with the UV colours of the Donas et al.\ (2007) ellipticals in the same apertures. We also plot the 49 ellipticals with $\log \sigma > 2.30$ km s$^{-1}$ from the Donas et al.\ (2007) sample in the top and bottom panels Figure \ref{fig:nonBCGs}. The UV colours of the Donas et al.\ (2007) ellipticals in the same mass range are also more scattered compared to the BCGs (with a standard deviation on the mean colour values of 0.42 compared to 0.33 for the BCGs). We again find that the ellipticals have bigger FUV--NUV colours than the BCGs (we find a mean FUV--NUV colour of 1.089$\pm$0.059, compared to 0.791$\pm$0.055 for the BCGs, and to 1.062$\pm$0.080 for the SB06 ellipticals). We find that the UV colour distributions of the two elliptical samples agree very well, whereas the distribution of the BCG sample is significantly different from both elliptical samples. 

It is known that BCGs have larger sizes and lower optical surface brightness than non-BCGs of the same mass (Von der Linden et al.\ 2007). Since we are measuring the UV fluxes in a radius scaled to the Kron radius (which itself is roughly equal to the half-light radius), it could be that we are measuring a slightly bigger FUV and NUV aperture for the BCGs than for the non-BCGs, which could influence the NUV--FUV colour measurements. Wang et al.\ (2010) simulated the differences arising in the NUV--r measurements of BCGs purely because of their structural differences. They found the difference in the NUV--r measurements (between the inner, R$_{50}$, and outer, R$_{90}$, galaxy apertures) to be 0.1 mag bigger for the BCGs than the non-BCGs, purely as a result of their structure. The corresponding difference in the FUV--NUV measurements can contribute to our result that the BCGs have a higher UV-upturn than the non-BCGs, but it can not fully explain the large differences in the distributions.

Wang et al.\ (2010) reported no difference between the FUV--NUV colours of BCGs and non-BCGs from SDSS data, however they cautioned that this might be ascribed to the lower quality of the FUV images used. It has to be noted that optically selected BCG samples have shown slightly different properties from X-ray selected samples (see Burke, Collins $\&$ Mann 2000; Wang et al.\ 2010), and the properties derived from the two different samples can usually be reconciled if luminosity-dependent evolution is taken into account. However, this does not explain why our result differs from Wang et al.\ (2010), as their FUV--NUV colour comparison was also based on an optically selected sample, and both samples define BCGs as the brightest galaxy closest to the centre of the cluster potential. Wang et al.\ (2010) also finds this null result for both their inner, R$_{50}$, and outer, R$_{90}$, galaxy apertures, thus the difference can not be as a result of different apertures.

We have also investigated how the scatter and range of the FUV--NUV colours in our sample compare to samples of BCGs with definite recent star formation. Most of the previous samples, such as the cooling flow hosted BCGs which showed evidence of recent star formation in Pipino et al.\ (2009), are at intermediate redshifts ($z \sim 0.2$ to 0.5) and should not be directly compared. We identify the nearby BCGs ($z < 0.1$) with blue cores in the optical data presented by Rafferty et al.\ (2008), and find five of these BCGs with reliable FUV--NUV measurements (calculated for the same apertures as used here). The FUV--NUV colours of these five galaxies range from --1.01 to 0.51 mag (average 0.06, with a standard deviation of 0.62). If we compare this value with the average (and standard deviation) found for the FUV--NUV colours of cooling flow hosted BCGs in our sample (0.68 $\pm$ 0.28), and for non-cooling flow hosted BCGs (0.80 $\pm$ 0.31), then we do not seem to see a strong influence of possible star formation connected to cooling flow clusters in our FUV--NUV measurements. Hence, the relative homogeneity of the FUV--NUV colours in the sample here might be used as evidence against sporadic episodes of star formation, since the range and scatter of FUV--NUV colours expected for such episodes is large compared to the scatter in our BCG sample. We do not analyse this further as our main focus is the variations in the UV-upturn (caused by HB stars).



\section{Origin of the UV-upturn: Testing model predictions}
\label{Model predictions}

Stars with high helium abundance can become UV-bright more easily, and therefore, it is believed that helium abundance can make the UV-upturn more pronounced (Dorman, O'Connell $\&$ Rood 1995; Yi, Demarque $\&$ Oemler 1997). 

It has been suggested that helium sedimentation occurs in the intracluster medium (ICM) of galaxy clusters (Abramopoulos, Chanan $\&$ Ku 1981; Gilfanov \& Syunyaev 1984; Qin \& Wu 2000; Chuzhoy \& Nusser 2003; Ettori \& Fabian 2006; Peng $\&$ Nagai 2009). Under the influence of gravity, the heavier helium nuclei accumulate at the centres of massive galaxy clusters and can produce significant amounts of He abundance ($\Delta Y > 0.25$) in the central regions of hot, massive clusters. If the UV-upturn is produced by extreme HB stars, the helium sedimentation process may help to make it more pronounced in those galaxies in the centres of the clusters. The UV-upturn phenomena should be most pronounced in high mass (or X-ray temperature), low redshift, and dynamically relaxed systems according to the Peng $\&$ Nagai (2009) model. The effect of helium sedimentation, and thus UV-upturn, should also be larger for BCGs in cool-core clusters than in non-cool core clusters.

\begin{figure}
   \centering
   \mbox{\subfigure{\includegraphics[scale=0.53]{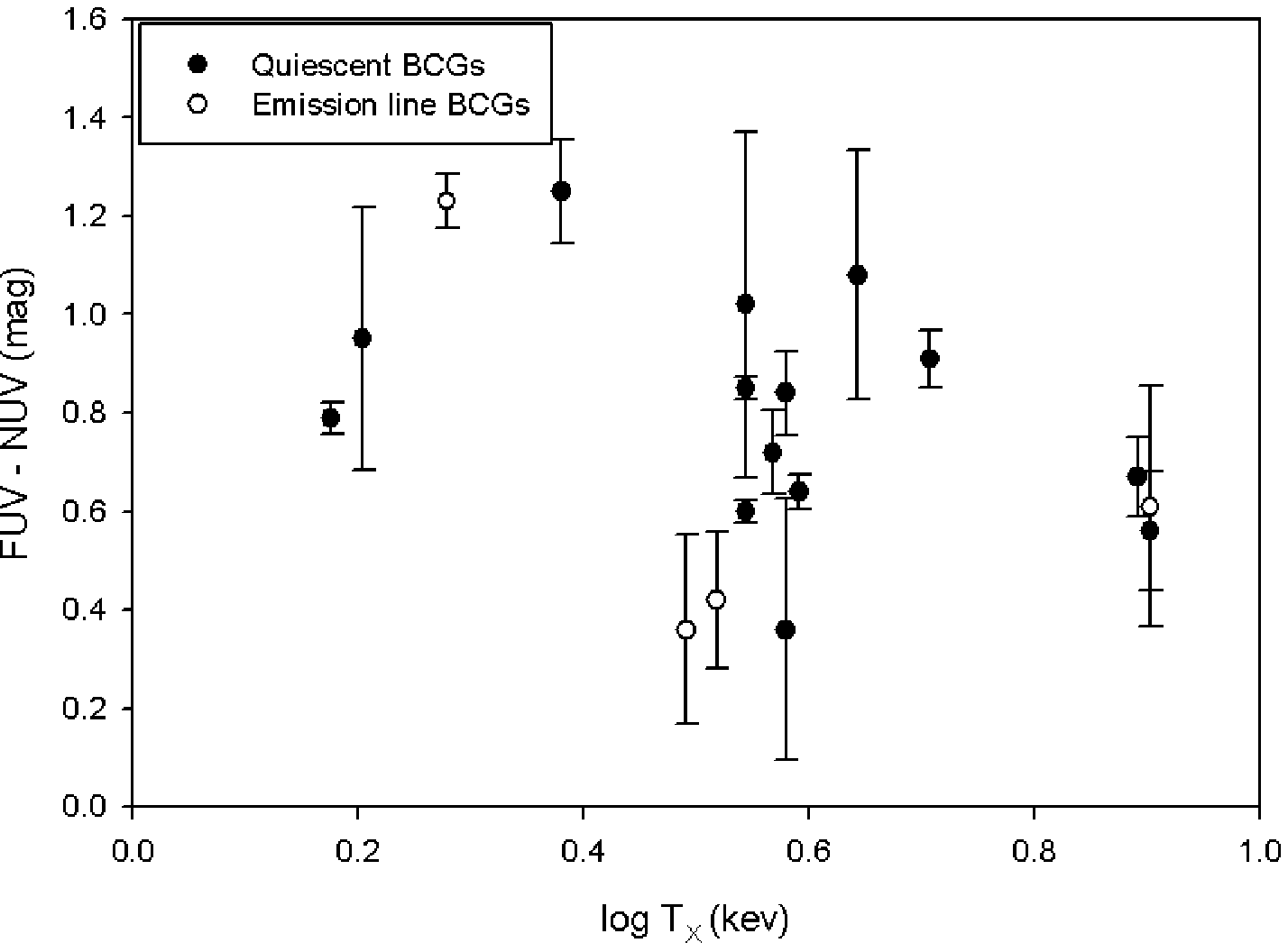}}}
   \mbox{\subfigure{\includegraphics[scale=0.53]{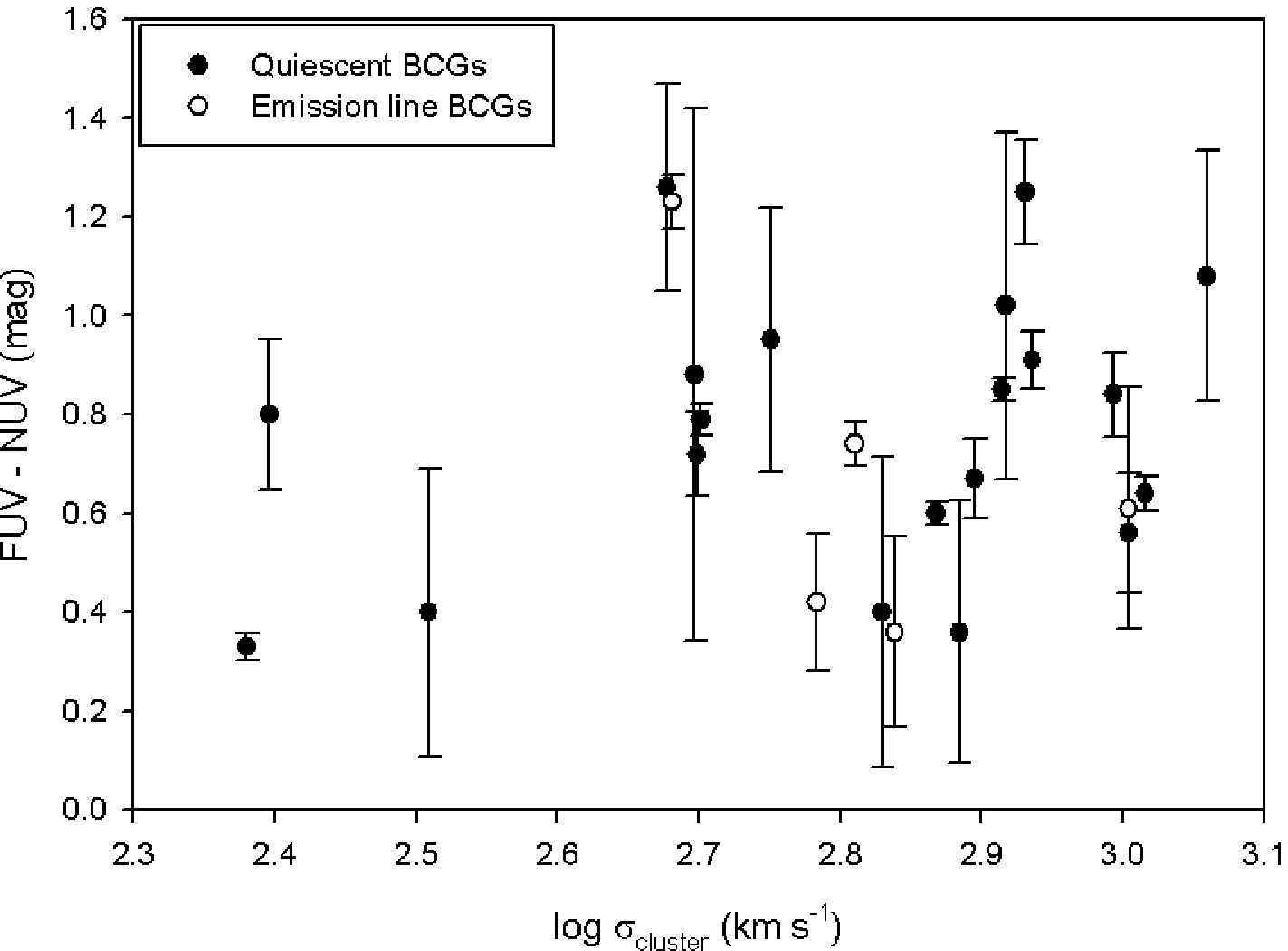}}}
   \caption{FUV--NUV colour against the X-ray temperature and velocity dispersion of the host clusters.}
   \label{fig:Tx}
\end{figure}

\begin{figure}
   \centering
   \includegraphics[scale=0.53]{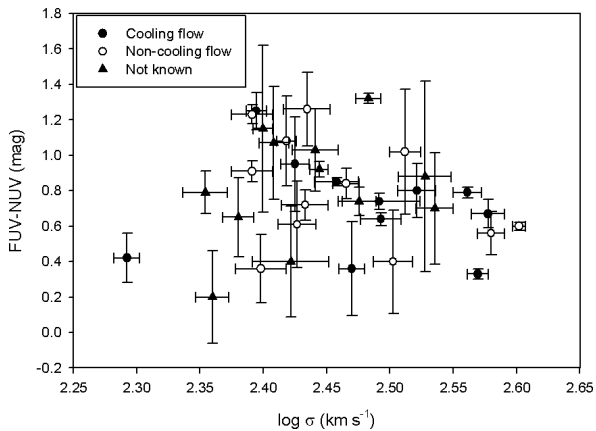}
   \caption{FUV--NUV colour against velocity dispersion for BCGs in cooling flows and non-cooling flows. Clusters with no cooling flow information available in the literature are also indicated.}
   \label{fig:Cooling}
\end{figure}

\begin{figure}
   \centering
   \includegraphics[scale=0.53]{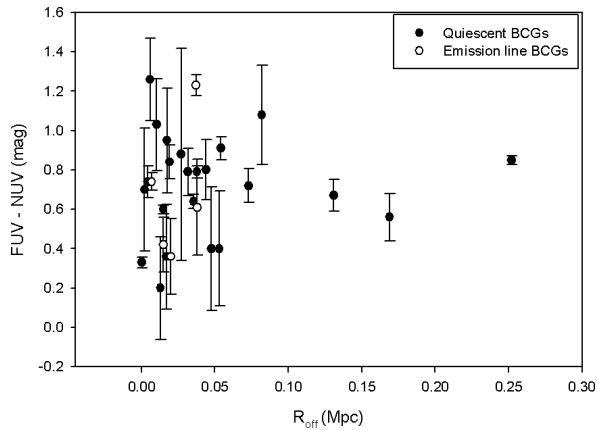}
   \caption{FUV--NUV colour against X-ray offset.}
   \label{fig:Roff}
\end{figure}

In particular, the Peng $\&$ Nagai (2009) model predicts:

\begin{enumerate}
 \item \textit{The UV-upturn correlates with X-ray temperature ($T_{\rm X}$), and the velocity dispersion of the host clusters ($\sigma_{cluster}$). }

We use $T_{\rm X}$ and $\sigma_{cluster}$ as proxies for the mass of the host cluster, and use the values from Table \ref{BigT} to plot Figure \ref{fig:Tx}. All the values are from spectra observed in the 0.1 -- 2.4 keV band with $ROSAT$, and using the same cosmology, namely the Einstein de--Sitter model of $H_{0}$ = 50 km s$^{-1}$ Mpc$^{-1}$, $\Omega_{\rm m}$ = 1 and $\Omega_{\Lambda}$ = 0. We do not convert to the Concordance model, as the X-ray temperature and cluster velocity dispersion is independent of the cosmological model assumed (White et al.\ 1997; Bohringer et al.\ 2004). All the values for X-ray temperature ($T_{\rm X}$ in keV) are from White et al.\ (1997). We find no clear evidence that the FUV--NUV colours correlate with $T_{\rm X}$ or $\sigma_{cluster}$ in Figure \ref{fig:Tx}.
 
\item \textit{The UV-upturn is more pronounced in BCGs than in non-BCGs, and in BCGs hosted by cooling flow clusters than hosted by non-cooling flow clusters.}

The former was plotted in Figure \ref{fig:nonBCGs}, and the latter is plotted in Figure \ref{fig:Cooling}. Figure \ref{fig:nonBCGs} showed significant differences in the distributions of the FUV--NUV colours of the BCG and non-BCG samples, with the non-BCGs having more pronounced differences in FUV--NUV (i.e. a lower UV-upturn and much more scatter, as discussed in Section \ref{Mass}). We find no difference between the FUV--NUV colours of the BCGs in cooling versus non-cooling flow clusters in Figure \ref{fig:Cooling}. 

It is interesting to note that Wang et al.\ (2010) found that among 21 X-ray selected BCGs (12 in cooling flows), those located in clusters with central cooling times of less than 1 Gyr are significantly bluer than those located in clusters where the central gas cooling times are long. However, this was based on NUV measurements. Unfortunately, in most cases we lack the necessary X-ray information to determine the cooling time.

\item \textit{The UV-upturn correlates with the dynamical state of the clusters. }

This is predicted as major mergers, for example, can destroy helium-rich cluster cores, producing some scatter in the UV-upturn among BCGs. Numerical simulations predict that the offset of the BCG from the peak of the cluster X-ray emission is an indication of how close the cluster is to the dynamical equilibrium state, and that this decreases as the cluster evolves (Katayama et al.\ 2003). Thus, the FUV--NUV colours were plotted against the offset between the X-ray peak of the cluster and the BCG ($R_{\rm off}$) in Figure \ref{fig:Roff}, where $R_{\rm off}$ is in Mpc and from Table \ref{BigT}. We find no evidence of a correlation between X-ray offset and FUV--NUV colour.

\end{enumerate}

The host clusters of our BCG sample cover a wide range in the parameters chosen to characterise the host clusters. The masses of the host clusters range from $\log  \sigma_{cluster}=2.3$ to 3.1 km s$^{-1}$, which includes groups to massive clusters. Similarly, the range of $T_{\rm X}$ and $R_{\rm off}$ is sufficient to detect possible trends (see Rafferty et al.\ 2008; Bildfell et al.\ 2008; Wang et al.\ 2010). In all cases, the errors on all the X-ray measurements are negligible compared to the scatter.

We also plotted Figures \ref{fig:Tx} and \ref{fig:Roff} distinguishing between cooling flow clusters, non-cooling flow clusters and clusters for which the information is not known, and we find no correlations.

In summary, we do find differences between the FUV--NUV colour distributions of normal ellipticals and BCGs in the same mass range, as predicted by the models. We do not find strong further evidence in our data to support the predictions made from the models of Peng $\&$ Nagai (2009) that the UV-upturn phenomena should be most pronounced in BCGs hosted by high mass, dynamically relaxed, cooling flow systems as a result of helium sedimentation. We also showed in Paper 2 that the BCG sample is not more uniform in age or metallicity than the SB06 sample in the same mass range. As such it is still an open question why we see differences, such as less scatter, between the FUV--NUV colours of the BCGs compared to the normal ellipticals. 

It is also interesting to note that Donahue et al.\ (2010) found no correlation between NUV colour and optical luminosity, X-ray temperature, redshift, or offset between X-ray centroid and X-ray peak.

As mentioned in Section \ref{Data}, we choose to characterise the UV-upturn as FUV--NUV colour based on the $GALEX$ measurements alone to minimise extinction and aperture effects and instrumental inhomogeneities. We note that this colour is not optimal for separating the effects of massive star formation from those of hot low-mass stars. To test whether this choice influences our results, we correlated the FUV--NUV colours against FUV--V colours for five BCGs in our sample which have homogeneous optical measurements in Sandage $\&$ Visvanathan (1978). We find a strong correlation between FUV--NUV and FUV--V within the errors (also see the correlation in figure 7 of Ree et al.\ (2007) for their nearby galaxies). We find that the scatter introduced in the FUV--V values by the large, uncertain aperture corrections is larger than the intrinsic scatter in FUV--NUV.

\section{Other correlations}
\label{Correlations}

A correlation between the UV-upturn and a property measurable from integrated light would not only help us to understand the origin of the UV-upturn but can also be used to utilise the UV-upturn as an age indicator (as discussed in Section \ref{Introduction}). 

For normal elliptical galaxies, the UV-upturn is reported to correlate with Mg, as well as with luminosity (Burstein et al.\ 1988; Donas et al.\ 2007; Gil de Paz et al.\ 2007). In particular, Gil de Paz et al.\ (2007) found elliptical galaxies with brighter $K$-band luminosities (i.e.\ more massive) to be redder in NUV--K but bluer in FUV--NUV than less massive ellipticals.

\begin{figure}
   \centering
   \mbox{\subfigure{\includegraphics[scale=0.53]{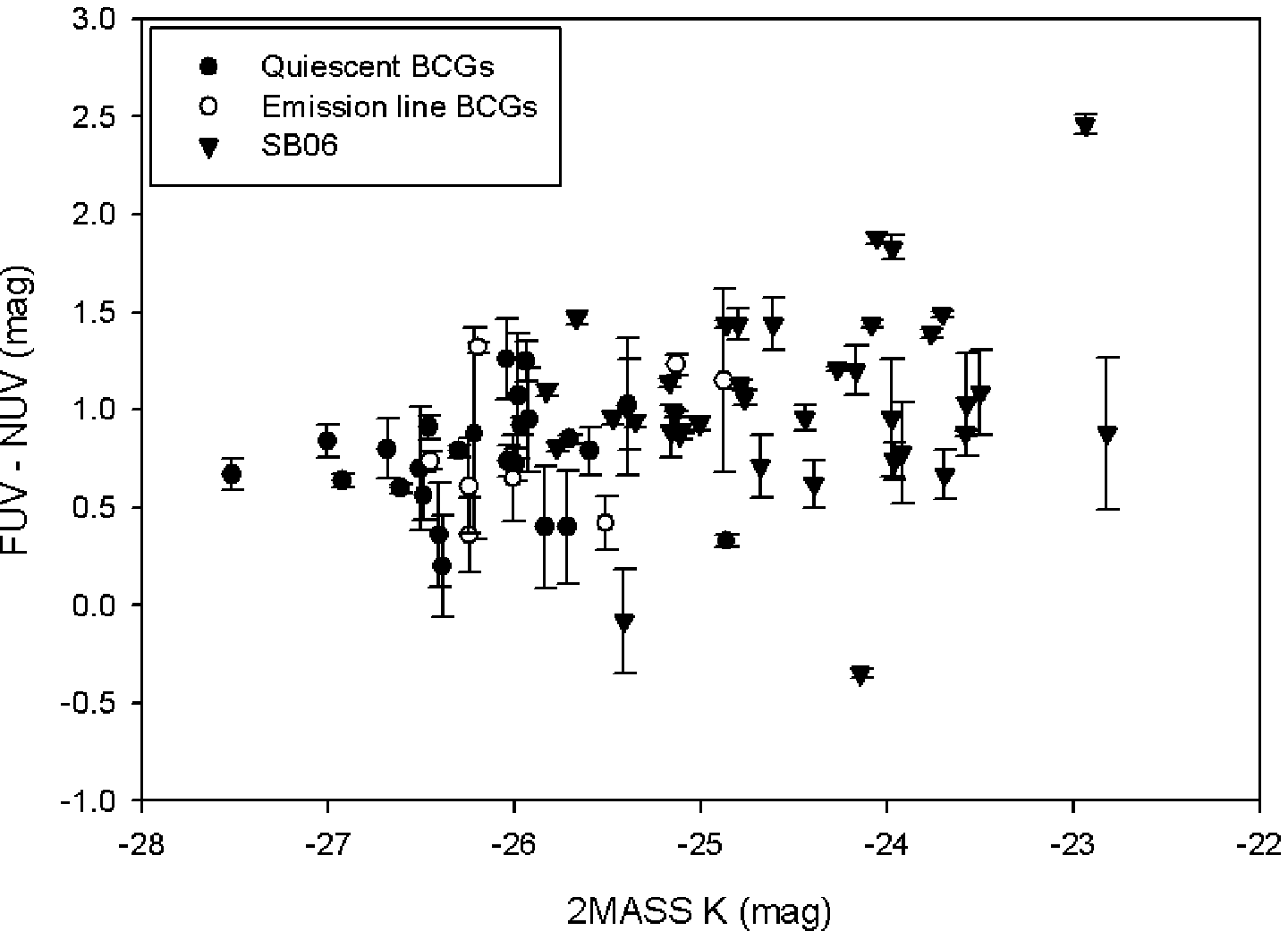}}}
   \mbox{\subfigure{\includegraphics[scale=0.53]{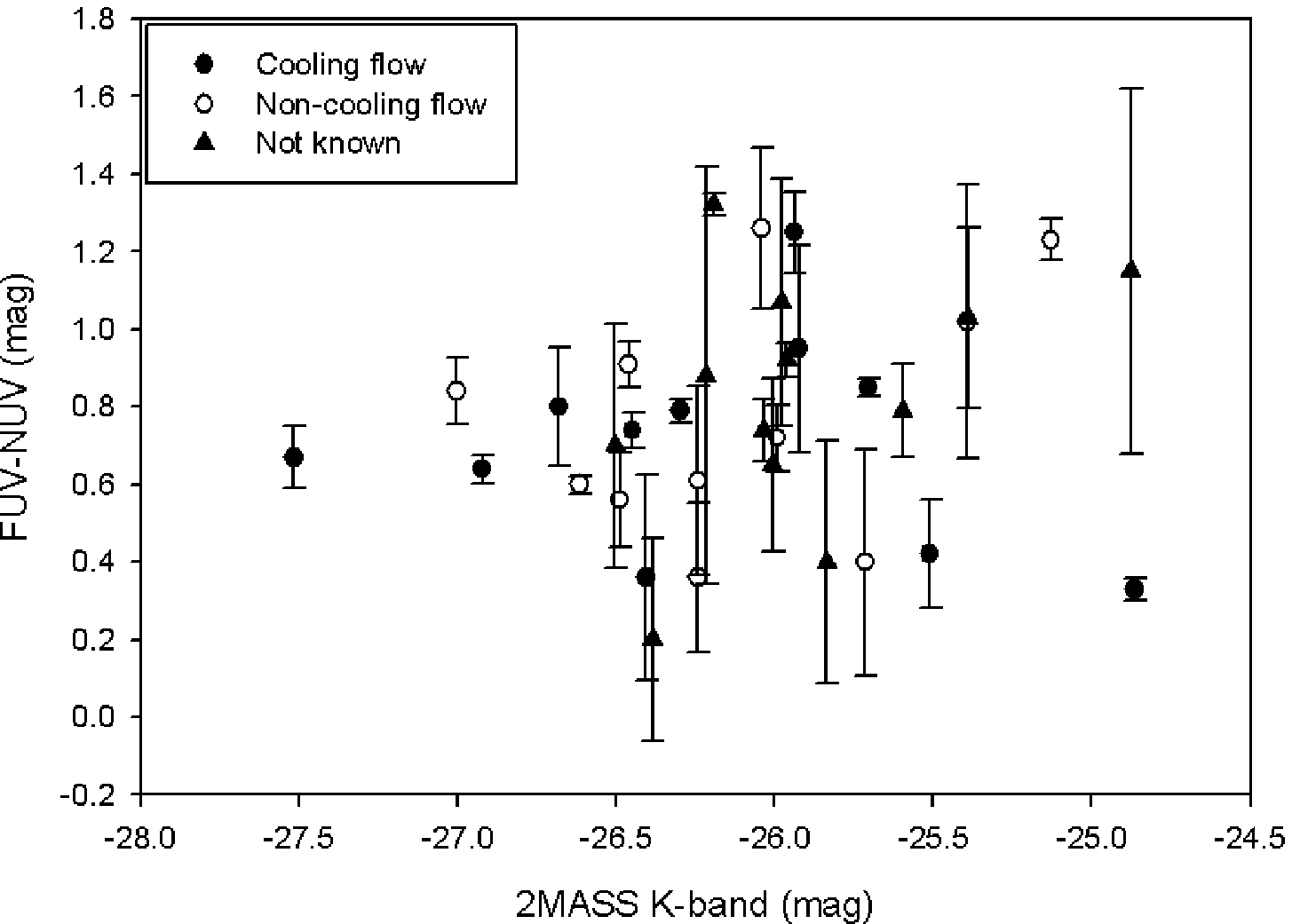}}}
   \caption{FUV--NUV colour against K-band luminosity. The top panel shows the SB06 sample of normal ellipticals for comparison. Clusters with no cooling flow information available in the literature are also indicated in the bottom panel.}
   \label{fig:Kmags}
\end{figure}

We plot the FUV--NUV colour against the 2MASS K-magnitudes in Figure \ref{fig:Kmags}, where we use different symbols to indicate emission line BCGs and quiescent BCGs, and BCGs hosted by cooling flow and non-cooling clusters, respectively. We also plot the 2MASS K-magnitudes of the SB06 sample of normal ellipticals in the top panel of Figure \ref{fig:Kmags}. The $K$-band is sensitive to the predominantly red populations in massive early-type galaxies and hence is a good measure of the total stellar mass. We do not correct the K-magnitudes for passive evolution. Using the Bruzual $\&$ Charlot (2003) stellar population synthesis code with the assumption that the galaxies are 10 Gyr old and formed in an instantaneous burst, this correction is only --0.2 mag for a galaxy at $z\sim0.054$ (in the $K$-band). Thus, it will not make a significant difference to whether or not a correlation is found.

We find no correlation with luminosity, however it is interesting to note that when the galaxies with very low FUV--NUV colour measurements are excluded (FUV--NUV $\lesssim$ 0.45 mag), then correlations are found between FUV--NUV colour and luminosity\footnote{This is clear in the bottom panel of Figure \ref{fig:Kmags}, whereas the top panel shows that the BCG outliers are still within the scatter of the normal elliptical galaxies.} as well as FUV--NUV colour and mass (Figures \ref{fig:nonBCGs} and \ref{fig:Cooling}). The low FUV--NUV colour measurements are not specific to lower/higher mass galaxies, galaxies with/without emission lines or galaxies hosted by cooling/non-cooling flow clusters. It is also not due to obvious systematic problems in the NUV and FUV measurements such as low exposure times for those galaxies. This is also the case for the outliers at the high end of FUV--NUV. 

\begin{figure*}
   \centering
  \includegraphics[scale=0.75]{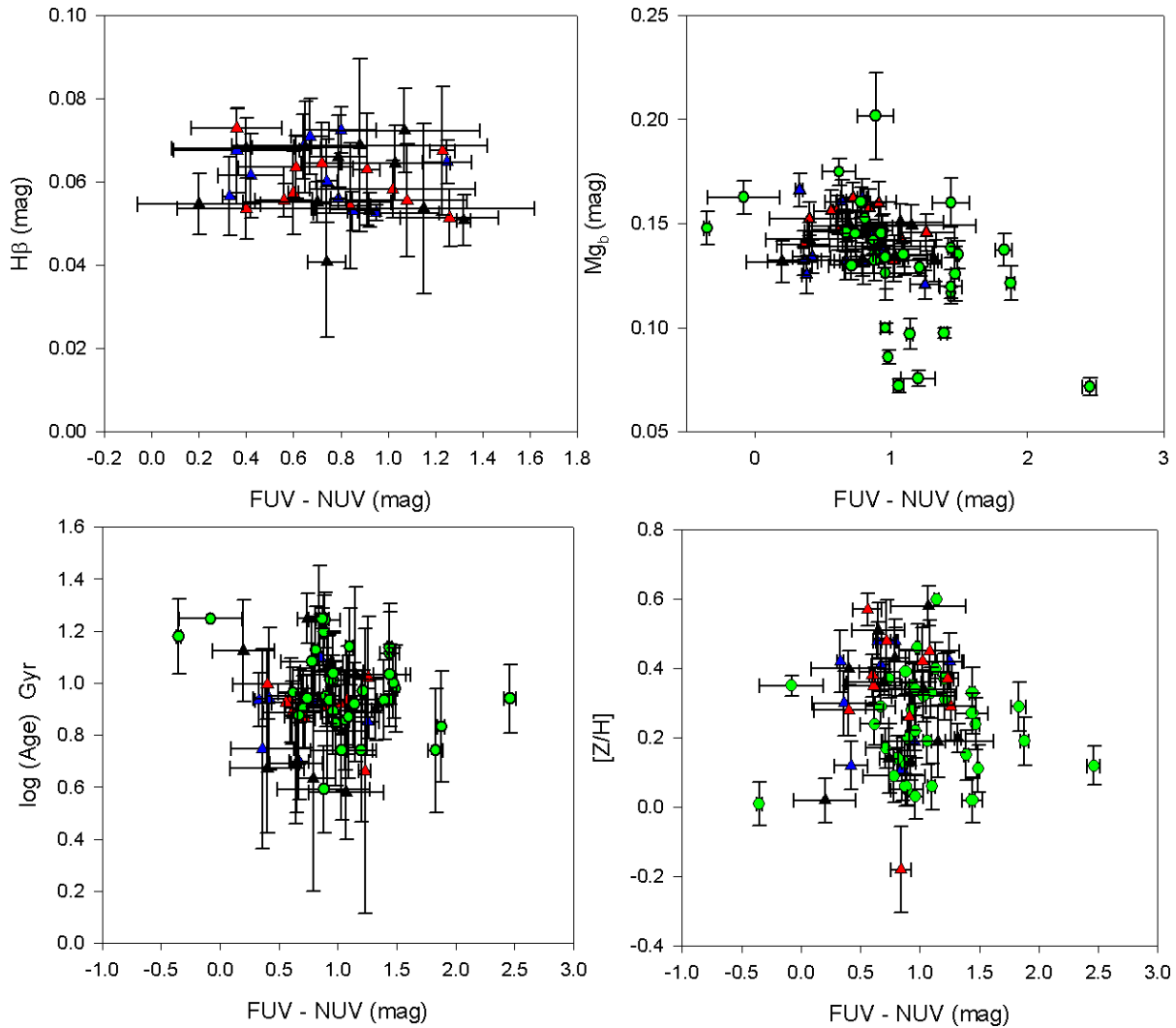}
   \caption{FUV--NUV colour against H$\beta$ and Mg$_{\rm b}$ index measurements, age and metallicity. The green symbols are for the SB06 elliptical sample in the same mass range. The blue triangles are BCGs hosted by cooling flow clusters, the red are BCGs hosted by non-cooling flow clusters, and the black are BCGs hosted by clusters of which the cooling flow status is not known from the literature.}
   \label{fig:Indices}
\end{figure*}

To analyse possible correlations between key absorption line indices, such as Mg$_{\rm b}$, and the FUV--NUV colours, we use the indices measured in the central apertures ($a_{\rm e}/8$)\footnote{The effective half-light radius was calculated as
$a_{\rm e} = \frac{r_{\rm e}(1-\epsilon)}{1-\epsilon \ \mid cos(\mid PA - MA \mid)\mid}$, with $\epsilon$ the ellipticity (data from NED), $r_{\rm e}$ the radius containing half the light of the galaxy (computed from the 2MASS $K$-band 20th magnitude arcsec$^{-2}$ isophotal radius), PA the slit position axis, and MA the major axis.}. We carefully calibrated these measurements to the widely used Lick/IDS system of absorption line indices (detailed in Paper 2), and plot the H$\beta$ and Mg$_{\rm b}$ indices against FUV--NUV colour in Figure \ref{fig:Indices}. We use Mg$_{\rm b}$ instead of Mg$_{2}$ here, since it is less affected by systematic errors for example flux calibration, and we re-measure the SB06 indices using the same procedure and plot them as reference. As the UV-upturn--Mg$_{\rm b}$ correlation is often interpreted as a metallicity effect on the UV flux, we also plot the FUV--NUV colours against the luminosity-weighted age and metallicity ([Z/H]), derived using the predictions of Thomas, Maraston $\&$ Bender (2003) and Thomas, Maraston $\&$ Korn (2004). To avoid artificial offsets between samples due to the use of different techniques to calculate the SSP-parameters, we recalculated these parameters for the sample of elliptical galaxies using exactly the same indices and method as used for our sample of BCGs. We find no significant correlation between H$\beta$ or Mg$_{\rm b}$ and FUV--NUV colour (using a $t$-test), and as in Figure \ref{fig:nonBCGs}, we find that the scatter in the FUV--NUV colour measurements for BCGs is much less in the plots in Figure \ref{fig:Indices} than for the ordinary ellipticals. As predicted by Yi $\&$ Yoon (2004), we do not find a correlation between $\alpha$-enhancement ([E/Fe]) and FUV--NUV colour (not shown here). In Paper 2, we found that BCGs are more metal-rich and have higher $\alpha-$abundance than the control sample of ellipticals, and here we have found that the BCGs have a higher UV-upturn, however the UV-upturn does not clearly correlate with [Z/H] or [E/Fe]. As mentioned in Section \ref{Data}, excluding emission line BCGs from these plots do not affect these conclusions. 

Donas et al.\ (2007) finds a tight anti-correlation between FUV--NUV and the Mg$_{2}$ index for nearby early-type galaxies, and Boselli et al.\ (2005) found a mild trend between FUV--NUV and Mg$_{2}$ in a sample of early-type galaxies in the Virgo cluster. On the other hand, Rich et al.\ (2005) detected no significant correlation between the FUV--r colour and the Mg$_{2}$ index or the velocity dispersion, using a large sample of SDSS early-type galaxies observed with $GALEX$. It is possible that this relation is diluted by lenticular galaxies, as shown in Donas et al.\ (2007). This correlation has not been previously investigated specifically for BCGs, and even though we do not find a statistically significant correlation between Mg$_{\rm b}$ (or Mg$_{2}$ -- not shown) and UV-upturn here, the properties are less scattered compared to our control sample of normal ellipticals.

\section{Summary}
\label{Discussion}

We compare the stellar population properties derived from optical spectroscopic data with $GALEX$ UV colour measurements for 36 nearby BCGs to provide a systematical comparison between the UV-upturn in BCGs with that of ordinary massive elliptical galaxies. We also test recent model predictions on the origin of the UV-upturn, specifically in central cluster galaxies. To do this we analyse possible correlations between the FUV--NUV colours and the internal properties of the galaxies (velocity dispersion, luminosity) and those of the host clusters (mass, distance from BCG to X-ray peak, presence of cooling flows).

We pay particular attention to the model of Peng $\&$ Nagai (2009). They pointed out that the helium sedimentation process can greatly increase the helium abundance in BCGs making them conductive to the formation of UV bright HB stars. It has long been suggested that helium sedimentation occurs in the ICM of galaxy clusters, and under the influence of gravity, the heavier nuclei in the H-He dominated ICM accumulate at the centres of the clusters. Thus, measuring the UV-upturn in BCGs compared to ordinary massive elliptical galaxies should provide a test of this scenario.

Our main conclusions are the following:

\begin{itemize}
\item We find significant differences between the UV-upturn distributions between BCGs and normal ellipticals in the same mass range, in that the BCGs have higher UV-upturns, and are less scattered than the normal ellipticals.

\item We do not find strong evidence in our data to support the predictions made from the models of Peng $\&$ Nagai (2009) that the UV-upturn phenomena should be most pronounced in BCGs hosted by high mass, dynamically relaxed, cooling flow systems as a result of helium sedimentation. 

\item We find no clear correlation between BCG velocity dispersion (mass) and the UV-upturn, but we find the UV-upturn of the BCG sample to be much less scattered with velocity dispersion than the non-BCGs over the same mass range. We also find no clear correlation between UV-upturn and luminosity, but the BCGs are less scattered than the normal ellipticals. 

\item We do not find the UV-upturn--Mg correlation reported for some samples of ordinary elliptical galaxies (Burstein et al.\ 1988; Donas et al.\ 2007), but we do find that the values are much less scattered than that of the normal elliptical sample that we use as reference. No correlations between the age, metallicity and $\alpha-$abundance of the BCGs with UV-upturn were found.
\end{itemize}

The UV-upturn is the most variable photometric feature associated with old stellar populations, yet it seems that the influence of the cluster environment is limited to enhancing the UV-upturn in BCGs compared to non-BCGs, and independent of the specific properties of the host cluster (also see Atlee et al.\ 2009). This absence of correlations with host cluster properties makes it difficult to invoke the helium sedimentation scenario as the primary mechanism responsible for the UV-upturn.

On both the theoretical and observational grounds, the lifetime UV outputs of these old HB stars are very sensitive to their physical properties (for example helium abundance). More remarkably, changes of only a few 0.01 solar masses in the mean envelope mass of an extreme HB population can significantly affect the UV spectrum of an elliptical galaxy (O'Connell 1999). If this interpretation is correct, then FUV observations become a uniquely delicate probe of the star formation and chemical enrichment histories of elliptical galaxies, but only once we understand the basic astrophysics of these advanced evolutionary phases so that we can explain the diversity in UV observations.

\section*{Acknowledgments}
We thank the anonymous referee for constructive comments which contributed to the improvement of this paper. SIL gratefully acknowledges funding from the South African SKA project, and thanks Russell Johnston for helpful discussions. PSB is supported by the Ministerio de Ciencia e Innovaci\'on (MICINN) of Spain through the  Ramon y Cajal programme. This work has been supported by the Programa Nacional de Astronom\'{\i}a y Astrof\'{\i}sica of the Spanish Ministry of Science and Innovation under the grant AYA2007-67752-C03-01.

\appendix

\section{BCG properties}

Photometric measurements of the BCGs, and X-ray properties and velocity dispersions of the host clusters for all 36 BCGs are listed in Table \ref{BigT}. The velocity dispersions and photometric measurements of the SB06 ellipticals used here are listed in Table \ref{BigE}.

\begin{table*}
\centering
\begin{scriptsize}
\begin{tabular}{l l c c c c c c c r c r c}
\hline Galaxy  & Cluster & BCG $\sigma$ & NUV & FUV & K-band & T$_{\rm X}$ & \multicolumn{2}{c}{Cooling Flow} & \multicolumn{2}{c}{$\sigma_{\rm cluster}$} & \multicolumn{2}{c}{R$_{\rm off}$} \\ 
               &         & (km s$^{-1}$) & (mag) & (mag) & (mag) & (keV) &  & ref & (km s$^{-1}$) & ref& (Mpc) & ref \\
\hline ESO202-043 & A S0479 & 256 & 19.28 $\pm$ 0.15 & 20.35 $\pm$ 0.30 & --25.98 & -- & -- & -- & -- & -- & -- & -- \\
 ESO303-005 & RBS521 & 276 & 19.64 $\pm$ 0.12 & 20.67 $\pm$ 0.22 & --25.39 & -- & -- & -- & -- & -- & 0.010 & cb \\
 ESO346-003 & A S1065 & 226 & 18.74 $\pm$ 0.05 & 19.53 $\pm$ 0.12 & --25.60 & -- & -- & -- & -- & -- & 0.032 & cr\\ 
 ESO444-046 & A3558 & 292 & 18.44 $\pm$ 0.03 & 19.28 $\pm$ 0.08 & --27.00 & 3.8 & X & e & 986 & w & 0.019 & e\\
 ESO488-027 & A0548 & 248 & 18.71 $\pm$ 0.03 & 19.96 $\pm$ 0.11 & --25.94 & 2.4 & $\checkmark$ & w & 853 & w & $\star$ & cb\\
 ESO541-013 & A0133 & 295 & 19.41 $\pm$ 0.15 & 19.77 $\pm$ 0.24 & --26.41 & 3.8 & $\checkmark$ & w,r & 767 & w & 0.017 & e\\
 ESO552-020 & CID 28 & 229 & 19.13 $\pm$ 0.14 & 19.33 $\pm$ 0.24 & --26.39 & -- & -- & -- & -- & -- & 0.013 & cb \\
 IC1101 & A2029 & 378 & 18.89 $\pm$ 0.04 & 19.56 $\pm$ 0.07 & --27.52 & 7.8 & $\checkmark$ & w,r & 786 & w & 0.131 & p\\
 IC1633 & A2877 & 400 & 17.59 $\pm$ 0.02 & 18.19 $\pm$ 0.01 & --26.61 & 3.5 & X & w & 738 & w & 0.015 & cb\\
 Leda094683 & A1809 & 332 & 20.55 $\pm$ 0.08 & 21.35 $\pm$ 0.14 & --26.68 & 3.7 & $\checkmark$ & w & 249 & w & 0.044 & p\\
 NGC0533 & A0189B & 299 & 17.43 $\pm$ 0.04 & 18.17 $\pm$ 0.07 & --26.04 & -- & -- & -- & -- & -- & 0.004 & cb \\
 NGC0541 & A0194 & 246 & 18.03 $\pm$ 0.02 & 19.26 $\pm$ 0.05 & --25.13 & 1.9 & X & w & 480 & w & 0.037 & cb \\
 NGC1399 & RBS454 & 371 & 15.23 $\pm$ 0.02 & 15.56 $\pm$ 0.02 & --24.86 & -- & $\checkmark$ & w & 240 & w & $<$0.001 & cb\\
 NGC1713 & CID 27 & 251 & 19.18 $\pm$ 0.17 & 20.33 $\pm$ 0.48 & --24.88 & -- & -- & -- & -- & -- & -- & -- \\
 NGC2832 & A0779 & 364 & 17.74 $\pm$ 0.01 & 18.53 $\pm$ 0.03 & --26.30 & 1.5 & $\checkmark$ & w & 503 & w & 0.038 & cl\\
 NGC3311 & A1060 & 196 & 17.60 $\pm$ 0.07 & 18.02 $\pm$ 0.13 & --25.51 & 3.3 & $\checkmark$ & w & 608 & w & 0.015 & pe\\
 NGC3842 & A1367 & 287 & 17.42 $\pm$ 0.01 & 18.27 $\pm$ 0.02 & --25.70 & 3.5 & $\checkmark$ & w & 822 & w & 0.252 & e\\
 NGC4839 & A1656 & 278 & 17.42 $\pm$ 0.02 & 18.34 $\pm$ 0.04 & --25.96 & -- & -- & -- & -- & -- & $\star$ & --\\
 NGC4874 & A1656 & 267 & 18.10 $\pm$ 0.08 & 18.71 $\pm$ 0.25 & --26.24 & 8.0 & X & w,e,g & 1010 & w & 0.038 & cb\\
 NGC4889 & A1656 & 380 & 17.52 $\pm$ 0.05 & 18.08 $\pm$ 0.12 & --26.49 & 8.0 & X & w,e,g & 1010 & w & 0.169 & e\\
 NGC6034 & A2151 & 325 & 19.88 $\pm$ 0.12 & 20.90 $\pm$ 0.36 & --25.39 & 3.5 & X & w,g & 827 & w & $\star$ & -- \\
 NGC6086 & A2162 & 318 & 19.43 $\pm$ 0.15 & 19.83 $\pm$ 0.27 & --25.72 & -- & X & g & 323 & s & 0.053 & cl\\
 NGC6160 & A2197 & 266 & 18.56 $\pm$ 0.06 & 19.51 $\pm$ 0.28 & --25.92 & 1.6 & $\checkmark$ & w,g & 564 & w & 0.017 & cc\\
 NGC6166 & A2199 & 310 & 17.44 $\pm$ 0.02 & 18.18 $\pm$ 0.04 & --26.45 & 4.7 & $\checkmark$ & w,e,g,r & 794 & w & 0.007 & e\\
 NGC6173 & A2197 & 304 & 18.22 $\pm$ 0.02 & 19.54 $\pm$ 0.02 & --26.19 & -- & -- & -- & -- & -- & $\star$ & --\\
 NGC6269 & AWM5 & 343 & 19.44 $\pm$ 0.16 & 20.14 $\pm$ 0.29 & --26.51 & -- & -- & -- & -- & -- & 0.002 & cc \\
 NGC7012 & A S0921 & 240 & 18.52 $\pm$ 0.08 & 19.17 $\pm$ 0.23 & --26.01 & -- & -- & -- & -- & -- & -- & -- \\
 NGC7597 & A2572 & 264 & 19.66 $\pm$ 0.12 & 20.06 $\pm$ 0.31 & --25.84 & -- & -- & -- & 676 & st & 0.048 & cc\\
 NGC7647 & A2589 & 271 & 18.66 $\pm$ 0.03 & 19.38 $\pm$ 0.08 & --25.99 & 3.7 & X & e & 500 & w & 0.073 & e\\
 NGC7649 & A2593 & 250 & 19.30 $\pm$ 0.12 & 19.66 $\pm$ 0.16 & --26.24 & 3.1 & X & w & 690 & w & 0.020 & cl\\
 NGC7768 & A2666 & 272 & 18.49 $\pm$ 0.06 & 19.75 $\pm$ 0.22 & --26.04 & 1.6 & X & w,g & 476 & w & 0.006 & cl \\
 PGC026269 & A0780 & 222 & 17.96 $\pm$ 0.01 & 18.34 $\pm$ 0.03 & -- & -- & $\checkmark$ & e & 641 & e,r & 0.015 & e\\
 PGC030223 & A0978 & 337 & 20.37 $\pm$ 0.28 & 21.25 $\pm$ 0.50 & --26.22 & -- & -- & -- & 498 & st & 0.027 & cb \\
 PGC072804 & A2670 & 311 & 19.87 $\pm$ 0.02 & 20.51 $\pm$ 0.03 & --26.92 & 3.9 & $\checkmark$ & w & 1038 & w & 0.035 & cb\\
 UGC00579 & A0119 & 246 & 18.99 $\pm$ 0.03 & 19.90 $\pm$ 0.05 & --26.46 & 5.1 & X & w,e & 863 & w & 0.054 & e\\ 
 UGC10143 & A2147 & 262 & 19.15 $\pm$ 0.08 & 20.23 $\pm$ 0.26 & -- & 4.4 & X & w,e,g & 1148 & w & 0.082 & e\\
\hline
\end{tabular}
\end{scriptsize}
\caption[X-ray Properties of the Host Clusters.]{\begin{footnotesize} Velocity dispersions and photometric measurements of the BCGs, and X-ray properties and velocity dispersions of the host clusters. The BCG $\sigma$ and $\sigma_{\rm cluster}$ values are in km s$^{-1}$ and the projected distance between the galaxy and the cluster X-ray peak (R$_{\rm off}$) is in Mpc. The $\star$ marks at R$_{\rm off}$ indicate the galaxy is not in the centre of the cluster but closer to a local maximum X-ray density, different from the X-ray coordinates given in the literature. The references are: w = White, Jones $\&$ Forman (1997); e = Edwards et al.\ (2007); g = Giovannini, Liuzzo $\&$ Giroletti (2008); r = Rafferty et al.\ (2006); cc = Calculated from Bohringer et al.\ (2000); cl = Calculated from Ledlow et al.\ (2003); cb = Calculated from Bohringer et al.\ (2004); cr = Calculated from Cruddace et al.\ (2002); st = Struble $\&$ Rood (1999); s = Struble $\&$ Rood (1991); p = Patel et al.\ (2006); pe = Perez et al.\ (1998). All the values for T$_{\rm X}$ are from White et al.\ (1997). \end{footnotesize}}
\label{BigT}
\end{table*} 

\begin{table}
\centering
\begin{scriptsize}
\begin{tabular}{l c c c c}
\hline Galaxy  & $\sigma$ & NUV & FUV & K-band  \\ 
               &   (km s$^{-1}$) & (mag) & (mag) & (mag) \\
\hline NGC2329 & 229 & 18.59 $\pm$ 0.06 & 19.48 $\pm$ 0.12 & --25.16 \\
NGC3115 & 284 & 14.67 $\pm$ 0.01 & 16.11 $\pm$ 0.01 & --24.09 \\
NGC3379 & 228 & 15.10 $\pm$ 0.01 & 16.49 $\pm$ 0.01 & --23.77 \\
NGC4261 & 303 & 15.81 $\pm$ 0.01 & 16.79 $\pm$ 0.01 & --25.14 \\
NGC4278 & 277 & 15.47 $\pm$ 0.01 & 16.35 $\pm$ 0.01 & --23.58 \\
NGC4365 & 258 & 15.06 $\pm$ 0.01 & 16.19 $\pm$ 0.01 & --24.79 \\
NGC4472 & 310 & 14.48 $\pm$ 0.01 & 15.44 $\pm$ 0.03 & --25.47 \\
NGC4594 & 259 & 13.82 $\pm$ 0.01 & 14.96 $\pm$ 0.01 & --25.17 \\
NGC4621 & 232 & 15.45 $\pm$ 0.01 & 16.66 $\pm$ 0.01 & --24.27 \\
NGC5796 & 274 & 18.01 $\pm$ 0.08 & 18.72 $\pm$ 0.14 & --24.68 \\
NGC5812 & 217 & 17.60 $\pm$ 0.04 & 18.80 $\pm$ 0.12 & --24.17 \\
NGC5813 & 257 & 16.33 $\pm$ 0.02 & 17.39 $\pm$ 0.03 & --24.77 \\
NGC5854 & 254 & 17.03 $\pm$ 0.01 & 19.49 $\pm$ 0.05 & --22.94 \\
NGC5846 & 248 & 16.07 $\pm$ 0.01 & 16.97 $\pm$ 0.02 & --25.10 \\
NGC6127 & 251 & 18.10 $\pm$ 0.03 & 18.72 $\pm$ 0.12 & --24.39 \\
NGC1453 & 337 & 17.44 $\pm$ 0.01 & 18.38 $\pm$ 0.03 & --25.36 \\
NGC2300 & 303 & 17.00 $\pm$ 0.02 & 17.93 $\pm$ 0.03 & --25.01 \\
NGC2693 & 322 & 17.36 $\pm$ 0.01 & 18.23 $\pm$ 0.02 & --25.11 \\
NGC0315 & 304 & 17.31 $\pm$ 0.01 & 18.41 $\pm$ 0.03 & --25.83 \\
NGC3608 & 203 & 16.24 $\pm$ 0.01 & 17.73 $\pm$ 0.01 & --23.70 \\
NGC3665 & 222 & 16.33 $\pm$ 0.02 & 17.77 $\pm$ 0.08 & --24.80 \\
NGC4374 & 301 & 14.75 $\pm$ 0.01 & 16.19 $\pm$ 0.01 & --24.86 \\
NGC4552 & 262 & 15.30 $\pm$ 0.01 & 14.95 $\pm$ 0.01 & --24.14 \\
NGC4649 & 368 & 14.34 $\pm$ 0.01 & 15.15 $\pm$ 0.01 & --25.77 \\
NGC0507 & 266 & 16.84 $\pm$ 0.01 & 18.31 $\pm$ 0.03 & --25.67 \\
NGC0584 & 224 & 16.20 $\pm$ 0.01 & 18.08 $\pm$ 0.03 & --24.06 \\
NGC0821 & 212 & 16.93 $\pm$ 0.01 & 18.76 $\pm$ 0.06 & --23.97 \\
NGC7052 & 201 & 18.89 $\pm$ 0.09 & 18.81 $\pm$ 0.25 & --25.41 \\
CGCG-159-43 & 231 & 19.92 $\pm$ 0.14 & 20.70 $\pm$ 0.22 & --23.92 \\
IC3959 & 226 & 19.84 $\pm$ 0.03 & 20.51 $\pm$ 0.12 & --23.69 \\
IC3973 & 256 & 19.88 $\pm$ 0.12 & 20.76 $\pm$ 0.37 & --22.82 \\
IC4051 & 299 & 19.41 $\pm$ 0.04 & 20.85 $\pm$ 0.13 & --24.61 \\
NGC4842A & 232 & 19.78 $\pm$ 0.03 & 20.52 $\pm$ 0.09 & --23.96 \\
NGC4864 & 221 & 19.61 $\pm$ 0.13 & 20.57 $\pm$ 0.27 & --23.98 \\
NGC4865 & 308 & 19.26 $\pm$ 0.09 & 20.35 $\pm$ 0.20 & --23.50 \\
NGC4867 & 247 & 19.63 $\pm$ 0.11 & 20.66 $\pm$ 0.24 & --23.57 \\
NGC4908 & 201 & 18.55 $\pm$ 0.03 & 19.51 $\pm$ 0.06 & --24.44 \\
\hline
\end{tabular}
\end{scriptsize}
\caption[]{\begin{footnotesize} The velocity dispersions and photometric measurements of the control sample of ellipticals. \end{footnotesize}}
\label{BigE}
\end{table} 

\bsp

\label{lastpage}

\end{document}